%% file: main.tex
\documentclass[sigconf]{acmart}

\AtBeginDocument{%
  }

\copyrightyear{2023} 
\acmYear{2023} 
\setcopyright{acmlicensed}\acmConference[CHI '23]{Proceedings of the 2023
CHI Conference on Human Factors in Computing Systems}{April 23--28,
2023}{Hamburg, Germany}
\acmBooktitle{Proceedings of the 2023 CHI Conference on Human Factors in
Computing Systems (CHI '23), April 23--28, 2023, Hamburg, Germany}
\acmPrice{15.00}
\acmDOI{10.1145/3544548.3581466}
\acmISBN{978-1-4503-9421-5/23/04}

\acmSubmissionID{4721}

\usepackage{colortbl}  
\usepackage{xcolor}
\usepackage{array}
\begin{document}

\newcommand{\techName}[1]{\textit{NFTDisk}}
\newcommand{\disk}[1]{\textit{Disk Module}}
\newcommand{\flow}[1]{\textit{Flow Module}}
\newcommand{\modify}[1]{\textcolor{black}{#1}}
\newcommand{\newmod}[1]{\textcolor{black}{#1}}
\newcommand{\yong}[1]{\textcolor{brown}{[Yong: #1]}}
\title{\techName{}: Visual Detection of Wash Trading in NFT Markets }

\author{Xiaolin Wen}
\orcid{1234-5678-9012}
\affiliation{%
  \institution{Sichuan University}
  \streetaddress{P.O. Box 1212}
  \city{Chengdu}
  \country{China}
}
\affiliation{
  \institution{Singapore Management University}
  \country{Singapore}
}
\email{wenxiaolin@stu.scu.edu.cn}

\author{Yong Wang}
\authornotemark[1]
\affiliation{%
  \institution{Singapore Management University}
  \country{Singapore}}
\email{yongwang@smu.edu.sg}

\author{Xuanwu Yue}
\affiliation{%
  \institution{Baiont Technology}
  \city{Nanjing}
  \country{China}}
\affiliation{%
  \institution{Sinovation Ventures AI Institute}
  \city{Beijin}
  \country{China}}
\email{xuanwu.yue@connect.ust.hk}

\author{Feida Zhu}
\affiliation{%
  \institution{Singapore Management University}
  \country{Singapore}}
\email{fdzhu@smu.edu.sg}

\author{Min Zhu}
\authornote{The corresponding authors.}
\affiliation{%
  \institution{Sichuan University}
  \city{Chengdu}
  \country{China}}
\email{zhumin@scu.edu.cn}








\begin{abstract}
With the growing popularity of Non-Fungible Tokens (NFT), a new type of digital assets, various fraudulent activities have appeared in NFT markets. Among them, \textit{wash trading} has become one of the most common frauds in NFT markets, which attempts to mislead investors by creating fake trading volumes.
Due to the sophisticated patterns of wash trading, only a subset of them can be detected by automatic algorithms, and manual inspection is usually required.
We propose \techName{}, a novel visualization for investors to identify wash trading activities in NFT markets, where two linked visualization modules are presented:
a radial visualization module with a disk metaphor to overview NFT transactions and a flow-based visualization module to reveal detailed NFT flows at multiple levels. 
We conduct two case studies and an in-depth user interview with 14 NFT investors to evaluate \techName{}.
The results demonstrate its effectiveness in exploring wash trading activities in NFT markets.
\end{abstract}





\maketitle

\input{texfiles/1_introduction}
\input{texfiles/2_relatedwork}
\input{texfiles/3_background}

\input{texfiles/4_method}

\input{texfiles/5_evaluation}
\input{texfiles/6_coclusion}

\bibliographystyle{ACM-Reference-Format}
\bibliography{reference}










\end{document}

%% file: texfiles/1_introduction.tex
\section{Introduction}
With the popularity of digital assets,\textit{ Non-Fungible Tokens (NFTs)} have captured increasing attention in recent years~\cite{wang2021non,ante2022non,bao2022non}.
As a special type of cryptocurrency, NFTs represent tradeable ownerships of digital assets (e.g., images, music, videos, and virtual creations), where the ownership is recorded in smart contracts on a blockchain~\cite{dowling2022non}.
The huge potential of NFT markets attracts a large number of investors, and at the same time, many fraudulent activities also follow~\cite{das2021understanding}.
\textit{Wash trading} is one of the most common fraudulent activities in NFT markets~\cite{tariq2022suspicious},
which attempts to mislead investors by creating fake trading volumes~\cite{victor2021detecting}. 
Wash trading is rampant in NFT markets due to the lack of effective regulation.
Das et al.~\cite{das2021understanding} have detected 9,393 instances of wash trading that generated more than 96 million dollars in trading volume across 5,297 NFT collections involving 17,821 users.
To keep investors from getting ripped off by wash trading, it is important and urgent to figure out how to find and analyze wash trading in NFT markets. 

Recent work has presented various strategies to detect wash trading activities automatically by recognizing specific network structures in NFT transaction networks~\cite{das2021understanding,von2022nft}.
To evade detection, wash traders are creating increasingly sophisticated transaction patterns by trading different NFTs among multiple anonymous addresses of their own, which makes it possible to avoid some features of wash trading detection in traditional financial markets, such as closed trading loops, and results in some false negatives for automatic algorithms~\cite{von2022nft}.
Hence, automatic methods can just identify a lower bound for the actual extent of wash trading and evaluate the risk of NFT markets at a relatively high level, and manual inspection is usually required~\cite{das2021understanding}.
Also, the genuine NFT investors can only get limited help directly from the labels provided by automatic algorithms and the high-level analysis reports, since it is difficult for them to understand what was going on in the historical trading of the NFT collection of their interests.
A new crucial concern is how to inform users of the NFT transactions on the blockchain in an intuitive way to support manually detecting the complicated wash trading behaviors and evaluating the impact by using their domain experience.

Visualization and visual analytics techniques have been widely used to enhance people's understanding of blockchain transaction data~\cite{tovanich2019visualization}.
Some existing work has leveraged visualization to detect fraudulent activities, such as money laundering~\cite{mcginn2016visualizing}.
However, these studies focus on visualizing transaction data for traditional cryptocurrencies, such as bitcoin~\cite{di2015bitconeview}. 
Unlike traditional cryptocurrencies, each NFT is distinct and non-fungible, making it critical to pay attention not only to the number of NFTs traded, but also to which NFTs they are.
Therefore, it is non-trivial to leverage visualizations to help users recognize and understand wash trading activities effectively.
The challenge is two-fold. First, NFTs are often released based on NFT collections.
Each NFT collection has a significant number of unique NFTs that are traded between different NFT holders.
This creates a large amount of transaction data.
How to design visualizations to support quick identification of suspicious transactions and addresses on such a large data scale requires further consideration.
Second, verifying wash trading activities requires observing detailed NFT transaction patterns across a set of addresses.
It is challenging to visualize NFT transactions to support analyzing NFT flows between groups of addresses at different levels, such as the whole group, individual addresses, or individual NFTs.

In this paper, we introduce \techName{}, an interactive visualization for detecting and analyzing wash trading behaviors in NFT markets, which contains two visualization modules: the \disk{} and the \flow{}.
To be specific, the \disk{} is a novel radial visual design with a disk metaphor to provide a quick overview of transactions in one NFT collection to help users recognize suspicious transactions and addresses, as well as evaluate their influence on price and trade volume. An address reordering strategy is integrated in the \disk{} to make addresses with more transactions closer together to reduce the visual clutter. 
The \flow{} is a novel flow-based design to visualize NFT transactions at multiple levels and allow users to further check the patterns of wash trading in detail. 
To evaluate the effectiveness and usability of \techName{}, we conducted two case studies and an in-depth user interview with 14 NFT investors.
The results demonstrate that \techName{} is useful and  effective for visually detecting and analyzing wash trading activities in NFT markets.
In summary, our contributions are listed as follows:

\begin{itemize}
    \item We formulate the design requirements for visually detecting and analyzing wash trading in NFT markets, together with NFT investment experts, and propose a top-down workflow to help investors filter out unwanted information from a large amount of NFT transactions and analyze wash trading activities effectively.
    \item We propose \techName{}, a novel interactive visualization, which consists of two novel visualization modules: a radial visualization module with a disk metaphor enabling users to quickly identify wash trading activities in NFT transactions, and a flow-based module to display detailed NFT flows between addresses at multiple levels.
    \item We conduct two case studies and in-depth user interviews with 14 genuine NFT investors to demonstrate the effectiveness and usability of \techName{}.
\end{itemize}









%% file: texfiles/2_relatedwork.tex
\section{Related Work}
In this section, we summarize the related work in the most
relevant fields, including wash trading analysis in NFT markets and visualization for blockchain fraud detection.
\subsection{Wash Trading Analysis in NFT Markets}
Wash trading is a well-known phenomenon in traditional financial markets and refers to the activity of repeatedly trading assets for the purpose of feeding misleading information to the market.
Cao et al.~\cite{cao2015detecting} were among the first to analyze wash trading by specifying trading patterns.
They conceptualized the basic structures of wash trading using a directed graph of traders and proposed a dynamic-programming algorithm to recognize suspicious transactions and traders.
Wash trading in cryptocurrency markets has recently received a lot of attention~\cite{cong2021crypto,eigelshoven2021cryptocurrency,fratrivc2022manipulation}.
Victor et al.~\cite{victor2021detecting} presented a heuristic method to detect wash trading and provided a systematic analysis of wash trading behavior on two popular decentralized exchanges on the Ethereum blockchain.
Cong et al.~\cite{cong2022detecting} introduced systematic tests exploiting robust statistical and behavioral trading patterns to detect wash trading on 29 cryptocurrency exchanges.
Chen et al.~\cite{chen2022cryptocurrency} presented a data mining-based method for detecting wash trading using both off-chain and on-chain data, and the results indicated that several cryptocurrency exchanges have obviously faked trading volume.

As a new type of cryptocurrency, NFTs are also suffering from the wash trading problem.
Several recent publications have analyzed wash trading in the NFT market.
Das et al.~\cite{das2021understanding} presented a systematic overview of the security issues in the NFT ecosystem, where they detected wash trading by recognizing strongly connected components in the NFT trade networks.
Tariq et al.~\cite{tariq2022suspicious} conducted multiple statistical tests and provided some evidence of anomalous price patterns and suspicious automated trading in a substantial portion of the NFT market.
Von et al.~\cite{von2022nft} detected and quantified wash trading behaviors in NFT markets and summarized some features of NFT wash trading. They utilized the Deep-First-Search algorithm to find closed cycles in the NFT transaction graphs and further detect suspicious trade paths with both a high trade speed and low price deviation.
CryptoSlam\footnote{https://blog.cryptoslam.io/}, an NFT analysis tool, detects the NFTs with wash trading by checking whether they have repeated holders and shows the label ``Wash Sale'' when displaying NFTs on their website.
In a word, existing work on NFT wash trading focuses on designing automatic detection strategies based on specific features of wash trading and then analyzing suspicious transactions at a relatively high level.

Wash trading in NFT markets is protean since users can hold multiple addresses anonymously, and wash traders are creating increasingly sophisticated patterns to evade existing heuristic detection.
Automatic detection can find a subset of wash trading behaviors~\cite{von2022nft}.
However, informing investors about the number of suspicious transactions or labeling NFTs is insufficient to provide investors with enough information to understand what is happening in NFT transactions.
For these reasons, we present an interactive visualization technique that allows investors to understand how the NFTs traded between addresses over time and analyze wash trading behavior in depth.

\subsection{Visualization for Blockchain Fraud Detection}
In recent years, various blockchain data visualization approaches have been proposed to help people understand data recorded on the blockchain in a more intuitive way~\cite{sundara2017study,zhong2020silkviser,tovanich2021miningvis,conforti2022cryptocomparator,ocheja2022visualization}.
Fraud detection is one of the most common tasks of blockchain data visualization~\cite{chawathe2018monitoring,song2018design}.
Most of the existing visualization tools for blockchain fraud detection rely on analyzing the \textit{cryptocurrency value flow} and the \textit{transaction network}~\cite{tovanich2019visualization}.

The \textit{value flow} depicts how the value of cryptocurrency is propagated across different entities over time, which can be used to differentiate some fraudulent activities.
For example, Di Battista et al.~\cite{di2015bitconeview} presented BitConeView, a system for the visual analysis of Bitcoin flows on the blockchain, to assist analysts in gaining an immediate understanding of when and how Bitcoins were mixing in a suspicious way.
Bistarelli et al.~\cite{bistarelli2017go} used node-link diagrams to represent the Bitcoin flows to help recognize frauds like money laundering.
Ahmed et al.~\cite{ahmed2018tendrils} used a node-link tree visualization to track the stolen coins and reveal the operational techniques of criminals.
Xia et al.~\cite{xia2020supoolvisor} presented a visual analytics system for monitoring the mining pool by visualizing the Bitcoin flows between miners.

The \textit{transaction networks} emphasize the relationship between various entities on the blockchain, such as addresses~\cite{isenberg2017exploring}, clusters~\cite{kinkeldey2017bitconduite,sun2022bitanalysis}, and exchanges~\cite{yue2018bitextract}.
McGinn et al.~\cite{mcginn2016visualizing} combined both the transaction and address graphs in one high-fidelity visualization of associations to overview Bitcoin transaction activity and support the discovery of unexpected high-frequency transaction patterns.
Sun et al.~\cite{sun2019bitvis} presented BitVis, an interactive system to visualize relationships between Bitcoin addresses and transactions, which can help users conveniently analyze transaction behaviors and track suspicious accounts.
Further, Sun et al.~\cite{sun2022bitanalysis} presented BitAnalysis, a system for effective interactive investigation of Bitcoin wallets (i.e., a cluster of bitcoin addresses), which can monitor interested wallets in real time and alert the user to new suspicious transactions at a wallet level.

Overall, previous work mostly focuses on some traditional cryptocurrencies such as Bitcoin~\cite{mcginn2016visualizing,bistarelli2017go} and Ether~\cite{bogner2017seeing,norvill2018visual}, but NFT is a new type of cryptocurrency with its own intrinsic features.
Each NFT is unique, and all historical transactions of each individual NFT can be tracked on the blockchain.
It is a big challenge to track unique NFTs within a large amount of transactions in the blockchain and detect fraudulent activities such as wash trading from the complex transaction network, which is the focus of this paper.




%% file: texfiles/3_background.tex
\section{Background}
In this section, we first provide the data description, and then introduce that how we formulate our research problem.
Finally, we summarize the design requirements of our method.
\subsection{Data Description}
\textit{Non-Fungible Token (NFT)} is a type of cryptocurrency that is derived by the smart contracts of Ethereum~\cite{wood2014ethereum}, which was first proposed in Ethereum Improvement Proposal (EIP)-721\footnote{https://eips.ethereum.org/EIPS/eip-721}.
Unlike traditional cryptocurrencies such as Bitcoin and Ether, NFT is unique and can not be exchanged like-for-like (i.e., non-fungible), making it suitable for identifying something in a unique way.
Therefore, NFT can prove the existence and ownership of digital assets, such as videos, images, or arts.
NFT can be traded in various \textit{NFT marketplaces}, such as OpenSea\footnote{https://opensea.io/} and LooksRare\footnote{https://looksrare.org/\label{looksrare}}, as well as transferred directly on the blockchain.
All previous holders of each NFT can be tracked since all NFT transactions are recorded on the blockchain, whether traded in the marketplaces or directly transferred. 

In NFT markets, NFTs are listed based on their \textit{NFT collections}. 
An \textit{NFT collection} is an assortment of digital assets released by an artist (or group of artists) containing a limited number of individual NFTs with unique \textit{token id}. Typically, most NFT collections consist of numerous NFTs that all conform to the same artistic style, with slight variations across each individual token.
Most investors trade NFTs based on their interests in NFT collections, and the level of suspicious activity varies significantly across NFT collections~\cite{von2022nft}.
For these reasons, our method detects and analyzes wash trading behaviors on a collection basis.

In this paper, we collected all transactions of NFTs in a given NFT collection from the Ethereum blockchain using EtherScan\footnote{https://etherscan.io/}, a block explorer for the Ethereum blockchain.
After pre-processing, such as filtering out irrelevant information, each \textit{transaction} used in our method includes:
\begin{itemize}
    \item \textit{Timestamp}: the timestamp that the transaction occurred at;
    \item \textit{TokenID}: a unique number representing the token id of an NFT in one collection;
    \item \textit{Value}: the number of Ether (the standard cryptocurrency on the Ethereum blockchain) contained in the transaction, which can be seen as the sale price of NFT and transferred from the buyer to the seller;
    \item \textit{Status}: a label of \textit{sale} or \textit{transfer} according to whether the \textit{Value} of the transaction is zero, where \textit{transfer} implies the NFT is transferred without any Ether while \textit{sale} represents that the NFT is sold at the price of \textit{Value}.
    \item \textit{FromAddress}: the address that sold or transferred NFT to another address;
    \item \textit{ToAddress}: the address that bought or got NFT from another address.
\end{itemize}



\subsection{Problem Formulation}
Financial Conduct Authority~\footnote{https://www.handbook.fca.org.uk/handbook/MAR/1/6.html?date=2016-03-07\label{definition}} defined wash trading as, \textit{``a sale or purchase of a qualifying investment where there is no change in beneficial interest or market risk, or where the transfer of beneficial interest or market risk is only between parties acting in concert or collusion, other than for legitimate reasons.''}
According to the definition, wash trading has two important features: a group of users colluded and pretended to create fake transactions (Feature one), and they traded NFTs without market risk (Feature two).

For Feature one, we need to know the characteristics of address collusion.
In NFT markets, NFTs are traded between various addresses, but wash traders can create an arbitrary number of addresses anonymously, which makes it more convenient to trade NFTs between their own addresses.
All transactions recorded on the blockchain really occurred and required a fee called \textit{Gas Fee}, and the fees are also needed for wash traders transferring money among fake addresses.
Therefore, the number of colluding addresses is usually limited~\cite{das2021understanding}.
Previous studies have reported that most wash trading in the cryptocurrency markets operated as a set of users heavily traded on only small number of NFTs~\cite{das2021understanding}.
Therefore, we need to find a limited number of suspicious addresses that heavily trade several NFTs between themselves.

For Feature two, trading without market risk means that there are no changes in the assets held by the colluding addresses after a series of transactions~\cite{cong2022detecting}, which can make many fake transactions seem like real deals without taking risk.
Due to the possibility of multiple addresses controlled by a same person, verify the changes on held assets should be made both at the individual address level and at the address group level.
Furthermore, since NFT is non-fungible, we should not only focus on the change in the NFTs held by suspicious addresses, like the wash trading detection for other assets, but also need to further check the transfer process of each unique NFT, called \textit{NFT flow} in this work, to confirm whether the wash trading behaviors really occurred.

In summary, our research problem is how to design a novel visualization to support the identification of addresses with a possibility of collusion, i.e., addresses heavily trading a small set of NFTs, as well as display the NFT flows between these addresses at multiple levels to check the features of wash trading.

\subsection{Design Requirements\label{sec:requirements}}
To collect the requirements that can guide our visualization design, we conducted a preliminary study with two experts (E1, E2). 
Both experts have conducted NFT investments for at least one year and have been using a few online analysis tools to guide their investments.
E1 has experience analyzing the NFT transactions on the blockchain by developing some programs, and E2 has participated in the release of some new NFT projects.
\modify{
We interviewed E1 and E2 one by one and each interview lasted about 40 minutes.
To confirm the importance of our study,
we first asked the experts about the current state of wash trading in the NFT market, their perspectives on wash trading, and the limitations of existing tools.
Then, we 
collected their feedback on their general workflow for their analysis of wash trading
and
the possible functions that they wanted our visual analytics system to support.
}
The specific design requirements are summarized as follows:
\begin{itemize}
\item[\textbf{R1}] \textbf{Analyze wash trading in the scope of NFT collection.}
Experts said that NFTs in the same collection usually have similar price trends, and investors are concerned about the prospects of the whole NFT collection, rather than specific NFTs.
Our method should support analyzing wash trading based on individual NFT collections.
\item[\textbf{R2}]\textbf{Recognize suspicious transactions and addresses from the overview.} Our method should provide a quick overview of the transaction history of the whole NFT collection, where the undesired information should be filtered out and the suspicious transactions and addresses should be easily distinguished in an effective way.
Experts pointed out that the more recent transactions are the more important for analyzing wash trading, so our method should support the time selection and make the recent transactions more clear in the overview.
Flexible interactions are also needed to be integrated in the overview to help users select a group of suspicious addresses for further investigation.
\item[\textbf{R3}]\textbf{Reveal wash trading features at multiple levels.} 
Our method should support analyzing the  changes in NFTs held by the selected suspicious addresses in detail for verifying whether the wash trading really occurred.
To be specific, it should be allowed to analyze the NFTs held by both the whole address group and individual addresses over time, as well as track the previous holders of each individual NFT.
\item[\textbf{R4}]\textbf{Display the detailed transaction patterns of wash trading.}
Experts said that investors care not only about whether there are any wash trading behaviors in the collection but also about how the colluding address group conducted the wash trading to avoid being cheated by them.
Our method should display correct and comprehensive information of NFT transactions, such as the \textit{Timestamp}, \textit{Status}, and \textit{TokenID}.
\item[\textbf{R5}]\textbf{Enable the evaluation of wash trading influence.} Experts said that it is necessary to provide evidence to evaluate the extent to which the given NFT collection has been impacted by wash trading. 
Our method should answer the following questions: when do the wash trading behaviors appear more frequently, how many NFTs and addresses are involved in wash trading, and what is the impact on the price and trading volume of the given NFT collection?
To this end, more information such as average price and trading volume should be included.

\end{itemize}

%% file: texfiles/4_method.tex
\section{NFTDisk}
\begin{figure*}[htb]
  \centering
  \includegraphics[width=1\linewidth]{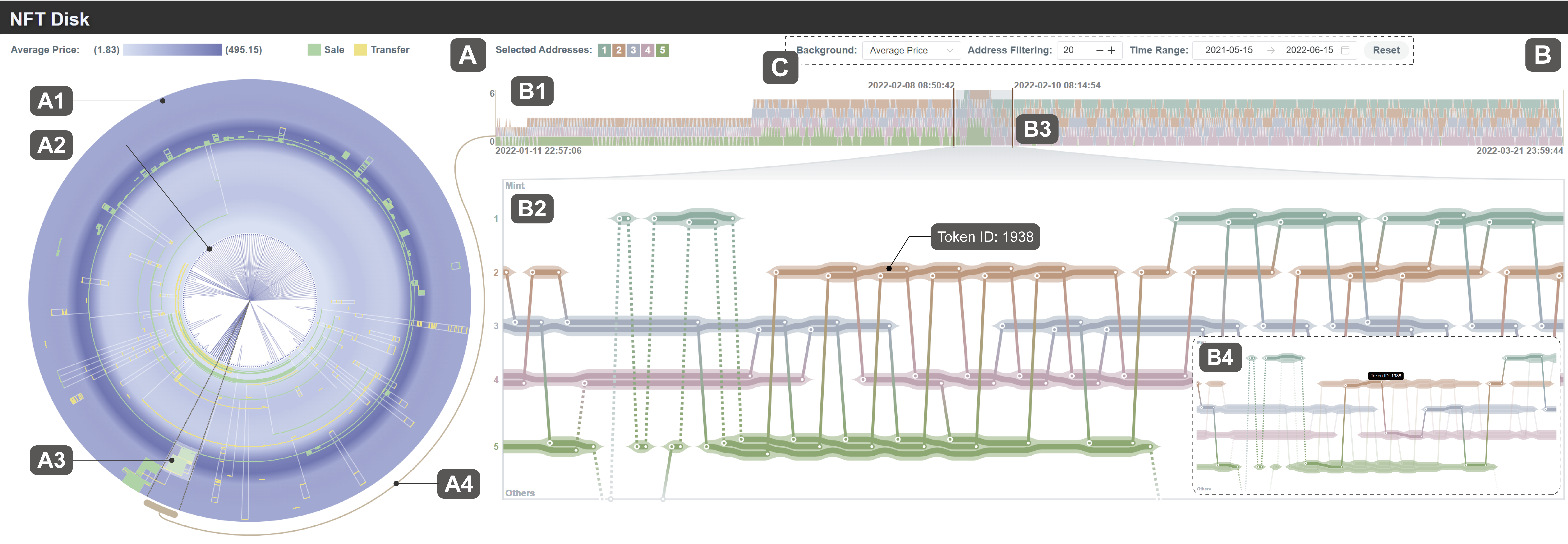}
  \caption{\label{interface}
            The \techName{} interface consists of the \disk{} (A), the \flow{} (B), and a tool bar (C) to provide some necessary interactions. The \disk{} contains an Outer Ring (A1) to overview NFT transactions and Inner Circle (A2) to display the suspicious score of address pairs. Users can brush an area (A3) in the \disk{} to select a group of addresses. In the \flow{}, a stacked area chart (B1) shows NFT flows at the group level, where users can brush a period (B3) to see detailed NFT flows in a flow-based chart (B2). Users can hover over the paths of each NFT to highlight them (B4).
           }
    \Description{Figure 1. fully described in the text.}
\end{figure*}

According to the design requirements above, we present \techName{}, an interactive visualization that supports detecting and analyzing wash trading behaviors at multiple levels, as shown in Figure~\ref{interface}.
\techName{} consists of two visualization modules: the \disk{} (Figure~\ref{interface}A) and the \flow{} (Figure~\ref{interface}B), and follows a drill-down workflow to help users filter out undesired information to concentrate on the wash trading behaviors.
First, the \disk{} visualizes the NFT transactions between addresses with heavy transactions as well as the average price and trade volume of the whole NFT collection (R1).
Users are allowed to find out the period and the group of addresses with suspicious transaction patterns from the overview (R2) and select them for further exploration in \flow{} by using a circular brush interaction (Figure~\ref{interface}A3).
The \flow{} displays the detailed NFT transfer process among addresses at the levels of address group, individual addresses, and individual NFTs, respectively (R3).
With the \flow{}, users can identify the features and transaction patterns of wash trading behaviors at multiple levels (R4).
After that, users can think about both the wash trading activities and the average price or trading volumes shown in the background of the \disk{}, and evaluate the extent to which the wash trading impacts this collection in a reasonable way (R5).
Following this workflow, users can easily detect and analyze the wash trading activities of a given NFT collection. 
The specific visual design and interactions are described in the following subsections.

\subsection{Disk Module}
The \disk{} is designed to provide an overview of NFT transactions on the blockchain, which is presented in a radial layout. 
It contains two parts: the \textit{Outer Ring} (Figure~\ref{interface}A1) to show the transaction details and the \textit{Inner circle} (Figure~\ref{interface}A2) to show the possibility of collusion of address pairs, where the Outer Ring encloses the Inner Circle.

\subsubsection{Outer Ring}
\begin{figure*}[htb]
  \centering
  \includegraphics[width=1\linewidth]{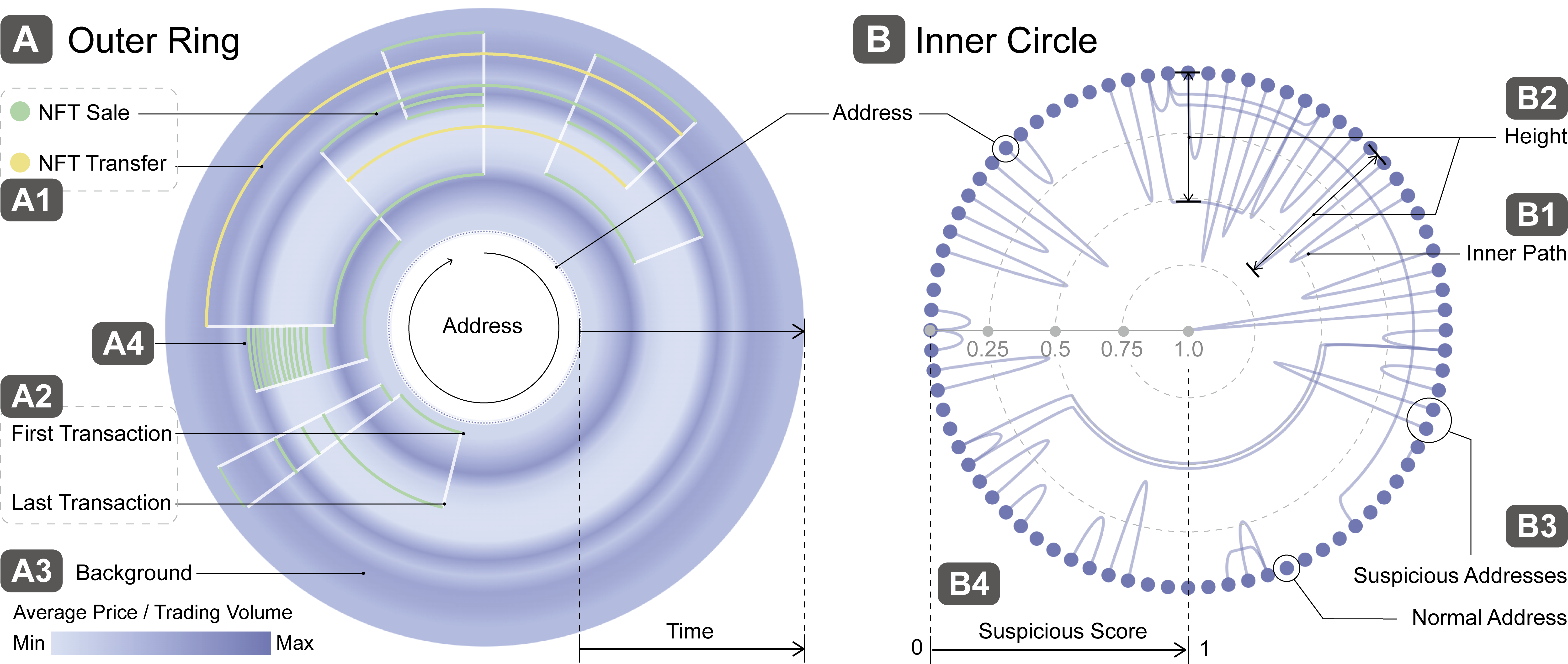}
  \caption{\label{disk}
            Visual encoding of \disk{}: In the Outer Ring (A), addresses are placed around the circle, while the timeline is from the inner circle to the outer circle. The green and yellow arcs (A1) represent the NFT sale and transfer between two addresses. The white line (A2) shows the first and last transactions at one address. 
            \modify{The gradient blue color of the background
            (A3) shows the average price or trading volume over time. The repeated transactions between two addresses can form a green block (A4).} In the Inner Circle (B), the inner path (B1) shows the suspicious scores (B4) of address pairs by their height (B2). The closer the inner path is to the center, the more suspicious the address pair is. (B3) shows some examples of suspicious addresses and 
a normal address. 
           }
    \Description{Figure 2. fully described in the text.}
\end{figure*}

In the Outer Ring, NFT transactions are displayed in the radial layout, which is extended from the circular Massive Sequence Views (MSV)~\cite{van2013dynamic}.
As shown in Figure~\ref{disk}A, the addresses are placed evenly around the circle, with each address encoded by a radius line at a specific angle.
\modify{
The time is represented by the concentric circles, with the smaller circles denoting an earlier timestamp and a larger circle representing a later timestamp.
The exact time range can be interactively configured by users.
NFT transactions are encoded by the arcs between the two radius lines at different angles that indicate the two involved addresses, as shown in Figure~\ref{disk}A1.}
The arcs have two colors encoding the status of transactions: green for ``Sale'' status and yellow for ``Transfer'' status.
We use the white lines along the radius (Figure~\ref{disk}A2) to link the first and last transaction of the addresses representing this radius.
The white lines can help users check whether the addresses are from new investors in this collection or not, because new investors with suspicious behaviors have a greater possibility of being the fake addresses created by wash traders.
Furthermore, a radial gradient blue background (Figure~\ref{disk}A3) is used to show the monthly average price or trading volume (R5), where the depth of the blue indicates the specific value and the change over time is represented by the inner circle to the outer circle.
\modify{
In the Outer Ring, we use absolute timestamps rather than the relative time sequence, as they help confirm the exact time when the wash trading occurred to assess the scale and impact.
But the use of absolute timestamps can result in 
overlapping arcs when several transactions occur at the same time or in a relatively short time period.
We deal with the overlapping arcs by allowing users to zoom into a shorter time span or brushing this period to observe details in the \flow{} (Figure~\ref{interface}B).
}
With the Outer Ring, users can overview the historical transaction patterns and analyze their influence on the average price or trading volume.

The reasons for the visual design are manifold. 
\modify{
First, according to R2, our design should make the colluding address groups with heavy transactions easier to distinguish.
The circular Massive Sequence Views (MSVs) have been shown able to visualize repeated patterns by utilizing Gestalt principles of closure and proximity~\cite{van2013dynamic}.
Inspired by it, we designed the Outer Ring (Figure~\ref{disk}A) to show the repeated transactions within address groups and imply wash trading behaviors. 
Specifically, the repeated transactions between two addresses can form an obvious green block in the Outer Ring (Figure~\ref{disk}A4).
Second, a circular layout can effectively reduce visual clutter by drawing the smaller of the two arcs between two addresses.
In Section~\ref{sec:requirements}, the experts pointed out that the more recent transactions are the more important for analyzing wash trading. 
Thus, our visualization is designed to put the more recent transactions closer to the outer circle, where there is more space, and it can make the recent transaction patterns more clear.
Third, to facilitate the impact analysis (R5), we added the gradient background to help compare the transactions, prices, and trade volumes based on the same timeline. 
}

\subsubsection{Inner Circle}
To further help users recognize suspicious addresses quickly, we present the Inner Circle in the center of the Outer Ring.
According to Das et al.~\cite{das2021understanding}, wash traders typically trade a small set of NFTs heavily, so it is suspicious when two addresses generate a large number of transactions but only involve a small number of NFTs.
To this end, we propose a measurement, \textit{Suspicious Score}, to indicate the possibility of colluding address pairs:
\begin{equation}
    S = 1 - \frac{N}{M},
\end{equation}
where $M$ is the number of transactions between the two addresses, and $N$ is the number of unique NFTs involved in these transactions.
The higher the suspicious score, the more likely the address pair is to collude.
If each transaction from a pair of addresses includes a different NFT, then their suspicious score is zero.

The Inner Circle is designed to display the suspicious score of address pairs in an intuitive way to make the suspicious address pairs easier to identify.
As shown in Figure~\ref{disk}B, each address is represented as a point around the circle, which is consistent with the layout in the \disk{}.
The points are connected by curves, called the \textit{inner paths} (Figure~\ref{disk}B1).
As shown in Figure~\ref{disk}B2, we define the height of the inner paths as the distance from the outer circle to its end on the side of the circle center, which represents the suspicious score of this address pair.
In other words, circles of different radius from the outer contour to the circle center represent suspicious scores from zero to one, as indicated by the dotted circle in Figure~\ref{disk}B4.
An inner path representing a pair of addresses always begins at one of them and travels through the dotted circle representing their suspicious score to the other, as shown in Figure~\ref{disk}B2.
It can also be regarded as the closer the inner path is to the center of the circle, the higher the suspicious score for this pair of addresses. 
If the suspicious score is zero, the inner path will not be drawn.
When users observe the inner circle, they can quickly identify the addresses with a high suspicious score according to the heights of the inner paths.
Figure~\ref{disk}B3 shows examples of suspicious addresses and a normal address, respectively.
\begin{figure*}[htb]
  \centering
  \includegraphics[width=1\linewidth]{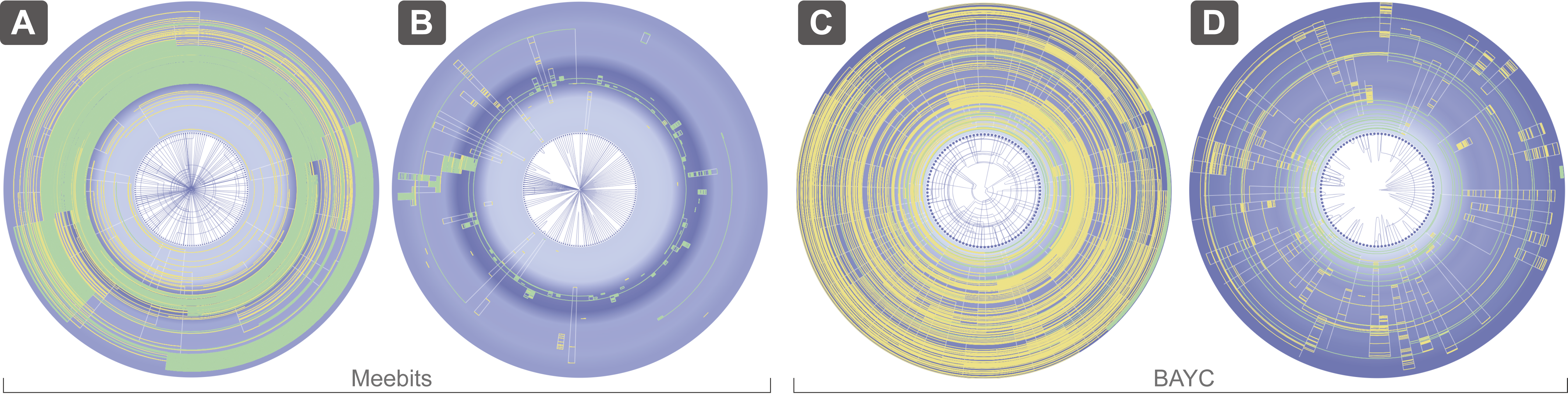}
  \caption{\label{reordering}
            The figures showcase the \disk{} with different address orders to show the effectiveness of our reordering strategy. (A) and (C) show the \disk{}s of two NFT collections named \textit{Meebits} and \textit{BAYC} with a random address order, while (B) and (D) show them after using our reordering strategy. Compared with (A) and (C), (B) and (D) have less visual clutter.
           }
    \Description{Figure 3. fully described in the text.}
\end{figure*}
\subsubsection{Reordering Strategy}
For the layout based on MSV, the node order is of vital importance, and a bad address order can seriously degrade the readability of the \disk{}, as shown in Figure~\ref{reordering}A and Figure~\ref{reordering}C.
Previous studies have proposed various reordering strategies to emphasize different features of MSV, such as reordering based on the node degree, edge length~\cite{van2013dynamic}, or neighbor relationships~\cite{linhares2017dynetvis}.
In this work, we propose a reordering strategy to make the addresses with more transactions closer.
Inspired by the reordering of rows or columns in matrix visualization~\cite{fekete2015reorder}, we regard address reordering as a \textit{Traveling Salesman Problem}~\cite{junger1995traveling} to get an order so that the address pairs with many transactions are as close to each other as possible.
Specifically, we convert the number of transactions between addresses into the distance between them, where two addresses with more transactions have a shorter distance.
Then, we use the \textit{optimal leaf ordering} algorithm~\cite{bar2001fast} to get an optimal address order to minimize the sum of distances between adjacent addresses.
The optimal leaf ordering algorithm will first compute hierarchical clustering according to the distance matrix between addresses to get a cluster tree, and then traverse the leaf nodes of the cluster tree according to the hierarchical structure to get the final order.
Figure~\ref{reordering} shows the \disk{} of two NFT collections named \textit{Meebits} and \textit{BAYC} using the random order or reordering by our strategy, which demonstrates that our strategy can effectively make the addresses with more transactions closer and make colluding groups look like blocks composed of many arcs.

\subsection{Flow Module}
\begin{figure*}[htb]
  \centering
  \includegraphics[width=1\linewidth]{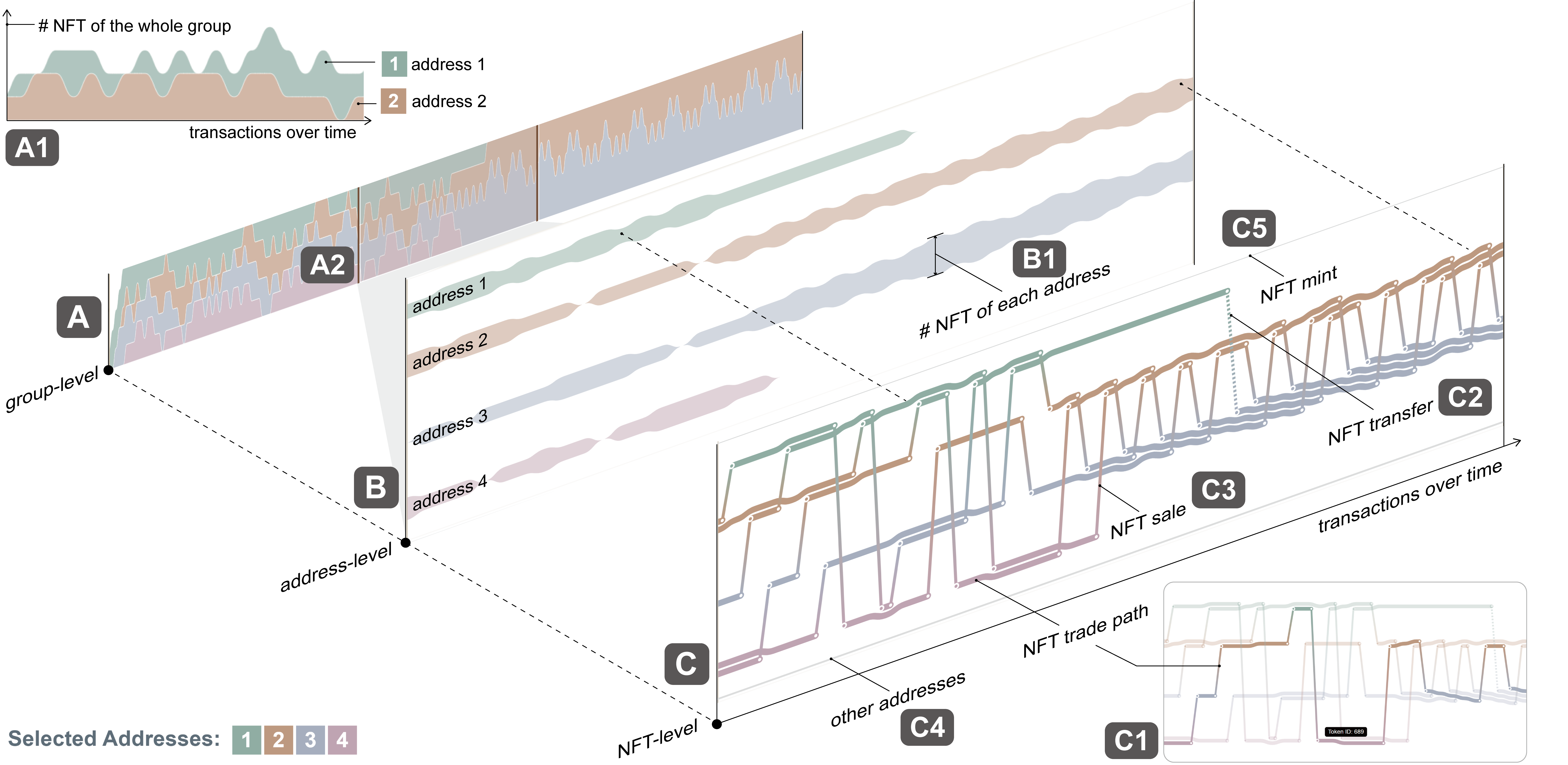}
  \caption{\label{flow}
            The multiple levels of \flow{}: The group level (A) shows the number of NFTs held by the whole selected address group using a stacked area chart (A1), where users are allowed to brush an area (A2) to check the details. The address level (B) shows the number of NFTs held by individual addresses, where each address is encoded with a flow graph (B1). The NFT level (C) superimposes on the address level, which demonstrates the trade paths (C1) of each NFT.
           }
    \Description{Figure 4. fully described in the text.}
\end{figure*}

The \flow{} (Figure~\ref{interface}B) is designed to visualize the detailed NFT flows at three different levels (R3): the group, address, and NFT levels.
The group and address levels are intended to display the NFTs held by the entire address group and the individual addresses over time, respectively, while the NFT level is intended to demonstrate the trading process of each individual NFT.
Specifically, the \flow{} contains two parts: a stacked area chart (Figure~\ref{interface}B1) and a flow-based chart (Figure~\ref{interface}B2), which cover the three levels above, as shown in Figure~\ref{flow}.
The specific visual designs are described as follows.


\subsubsection{Stacked Area Chart}
In the \flow{}, a stacked area chart is used to demonstrate the number of NFTs held by the selected addresses over time at the group level, as shown in Figure~\ref{flow}A.
The X-axis is a timeline, indicating the relative time sequence of the transactions rather than the absolute timestamp.
The height of the whole area (Y-axis) represents the number of NFTs held by this group after each transaction over time, as shown in Figure~\ref{flow}A1.
We use a categorical color scheme to represent different addresses, and the whole area is divided into parts in different colors by the individual addresses to show the number of each address in this group.
Users can identify temporal changes in the number of NFTs held by the selected group and roughly deduce the number of NFTs held by each address from the heights of different colored areas using the stacked area chart.
If the height of the stacked area chart remains constant over time (Figure~\ref{flow}A2), it indicates that NFTs flow only between this group of addresses, which is probably wash trading activity.
Users are also allowed to brush a period in the stacked area chart to further analyze the detailed NFT flows in the flow-based chart below.

\subsubsection{Flow-based Chart}
The flow-based chart (Figure~\ref{interface}B2) is the core of the \flow{}, which is designed to visualize the number of NFTs over time at the address level as well as show the detailed transaction patterns at the NFT level.
As shown in Figure~\ref{flow}B, the Y-axis places the addresses that have transactions in the selected period, and the X-axis represents the brushed time period in the stacked area chart.
The order of addresses is the same as the order they are stacked in the stacked area chart as well as their order in the \disk{}.
The NFTs held by each address are represented by a horizontal flow graph in the color representing this address, where the height of the flow graph represents the number of NFTs held by this address, as shown in Figure~\ref{flow}B1. 
Users can identify the temporal evolution of NFTs held by each address by the height change of the flow graph along the X-axis.

For each individual NFT, we use a path to visualize its transition process among its previous holders, which is called an NFT trade path (Figure~\ref{flow}C1) in this work.
If one address is holding an NFT in a time period, the path of this NFT will be packed in the flow chart of this address in this time period, where the path is in the same color as the holder (Figure~\ref{flow}C).
When the NFT is transferred or sold to another address, the path will link from the \textit{fromAddress} to the \textit{toAddress} with a dotted line (Figure~\ref{flow}C2) or a solid line (Figure~\ref{flow}C3), respectively.
In addition, if one party in the transaction is an address outside the selected address group, the path will link to the bottom border (Figure~\ref{flow}C4).
And if the NFT is got from the owner of the NFT collection, called \textit{Mint} in NFT markets, the path will link from the top border (Figure~\ref{flow}C5). 
The segment of the path that represents transactions is filled in a linear gradient color from the color of the \textit{fromAddress} to the color of the \textit{toAddress}.
When an address holds multiple NFTs, the paths of these NFTs are placed side by side and just fill the flow graph of this address.
To reduce the crossing of paths, the NFT is inserted from the side of \textit{fromAddress} (above or below this address) when getting a new NFT.
In this way, users can find out not only how many NFTs these addresses held by the contour of these paths and flow graphs in background, but also the detailed NFT transaction patterns by these trade paths.

\subsection{Interactions\label{sec:interaction}}
\techName{} enables rich interactions to allow users to smoothly explore the wash trading behaviors in NFT transactions.

\textbf{Visualization Configuration.} In Figure~\ref{interface}C, \techName{} allows users to configure the visualization to filter out some undesired information.
The \textit{Background Selector} allows users to select between monthly average price or trading volume to be shown in the background of the \disk{}.
The \textit{Time Selector} allows users to set the time range of their interest.
The \textit{Address Filter} allows users to filter the address pairs by the number of transactions between them to select suspicious addresses that are possible to collude with each other. 
Users set the threshold according to their experience, and then the address pairs with more than the threshold will be visualized in the \disk{}.

\textbf{Multi-level Brush.} 
\techName{} provides two types of brush interactions to help users drill down to the details of their interests.
Firstly, a circular brush is used in the \disk{} (Figure~\ref{interface}A3), where users can brush an arc area to select a group of suspicious addresses as well as a time period for further analysis in the \flow{}. 
For an arc area, the inner radius and outer radius determine the time range, while the start angle and end angle determine a set of addresses between the two angles.
After brushing, a curve is drawn to link the selected area to the flow module, as shown in Figure~\ref{interface}A4.
The second brush is integrated in \flow{} at the group level (i.e., the stack area chart) to allow users to select a time period, and then the detailed transactions of this period will be displayed below.


%% file: texfiles/5_evaluation.tex
\section{Case Studies\label{sec:case}}
We describe two case studies in this section to demonstrate the effectiveness of \techName{} in analyzing wash trading activities in real NFT collections.
The two cases were conducted by two users (U1, U2) during the user interview that will be introduced in Section~\ref{sec:interview}, where users were asked to learn \techName{} and use it to analyze NFT collections.
To demonstrate the generalization of \techName{}, four popular NFT collections with different features were used in our study, and all historical transactions of these collections from their launch time to June 30, 2022, were collected.
These collections are described as follows:

\textbf{\textit{Bored Ape Yacht Club (BAYC)}}\footnote{https://boredapeyachtclub.com/} is a collection of 10,000 Bored Ape NFTs—unique digital collectibles that was launched on April 15, 2021.
It is one of the most popular and expensive NFT collections in NFT markets, held by 6,460 unique owners, and has a total market cap of more than one billion dollars as of June 30, 2022.

\textbf{\textit{Meebits}}\footnote{https://meebits.app/} is an NFT collection of 20,000 unique 3D voxel characters, created by a custom generative algorithm, then registered on the Ethereum blockchain.
It was launched on May 4, 2021, was held by 6,572 unique owners, and had a total market cap of more than 110 million dollars as of June 30, 2022.

\textbf{\textit{Azuki}}\footnote{https://www.azuki.com/} is a collection of 10,000 generative avatar NFTs launched on January 12, 2022. 
These avatars have unique characteristics based on anime-themed drawings, which have caught the attention of NFT enthusiasts around the world.
As of June 30, 2022, it was held by 5,062 unique owners and had a total market cap of more than 118 million dollars.

\textbf{\textit{Loot}}\footnote{https://lootnft.io/} is a collection of 7,779 NFTs that is randomized adventurer gear generated and stored on blockchain. It is a typical collection based on the NFT gaming platform and was launched on August 27, 2021.
As of June 30, 2022, it was held by 2,567 unique owners, and had a total market cap of more than 18 million dollars.

\subsection{Case 1: In-depth Investigating of Wash Trading within an NFT Collection}
\begin{figure*}[htb]
  \centering
  \includegraphics[width=1\linewidth]{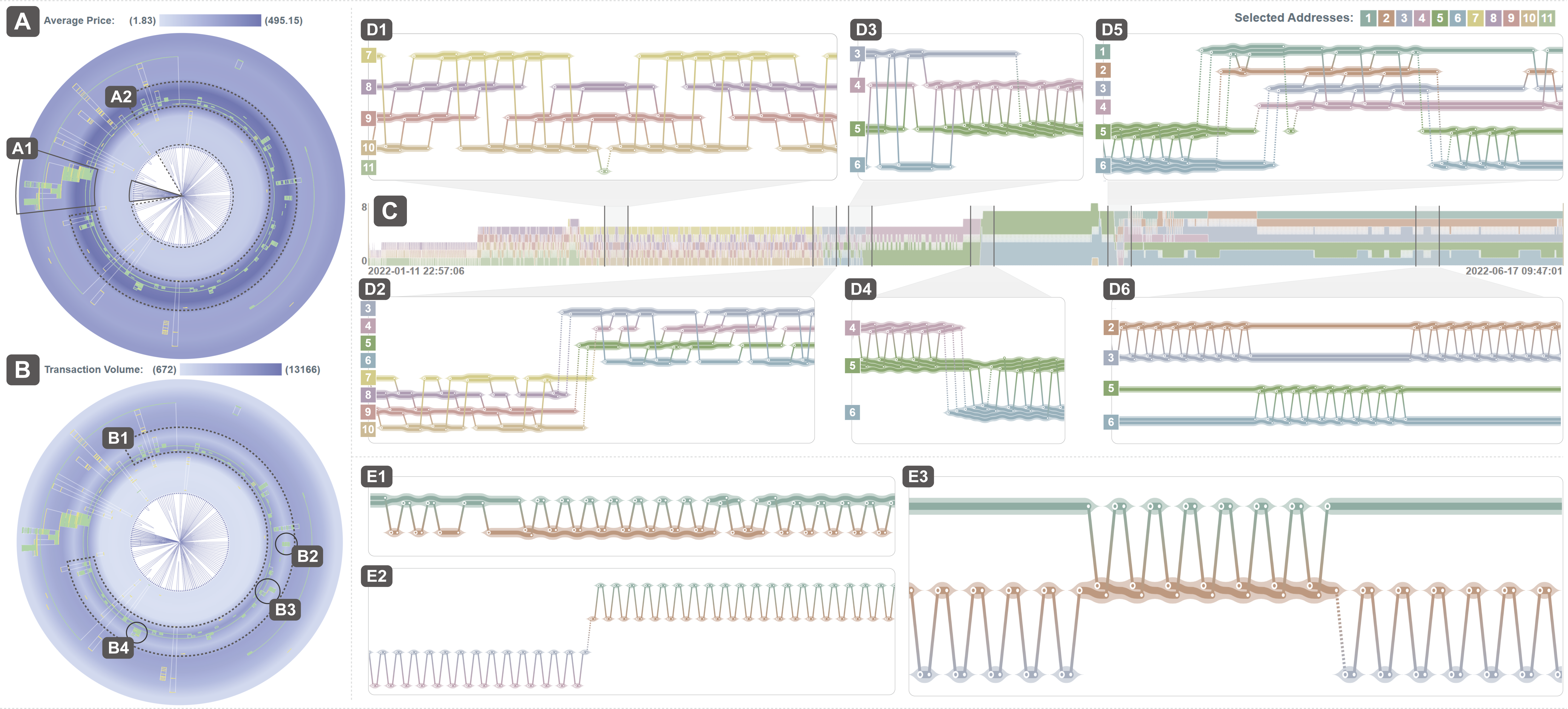}
  \caption{\label{case-1}\techName{} demonstrates various wash trading activities in the \textit{Meebits} collection.
            (A) and (B) show the \disk{} of \textit{Meebits} with the background representing the average price and trading volume respectively. In \disk{}, a big green block (A1) and many small green blocks (A2) are obvious. (C) shows the \flow{} at the group level after brushing the green block in (A1).
            (D1-D6) demonstrate the flow-based chart after brushing six different time periods in (C). (E1-E3) show the \flow{} after brushing areas in (B2-B4), respectively.
           }
    \Description{Figure 5. fully described in the text.}
\end{figure*}

U1 is the builder of an NFT community and often analyzes NFT transactions to help investors in the community.
When analyzing the \textit{Meebits}, one of the collections that U1 analyzed in the user interview, he first set the time range from October 1, 2021, to June 30, 2022, and set the \textit{Address Filter} to 20 according to his experience.
Then, the \disk{} provided an overview of NFT transactions during this period, where the gradient blue background represented the average price, as shown in Figure~\ref{case-1}A.
In the Inner Circle, he found that most of the inner paths were so close to the circle center, i.e., with a high height, which meant that addresses connected by these inner paths had a high suspicious score of collusion.
In the Outer Ring, he found that there were a lot of obvious shapes consisting of many consecutive arcs, of which a large green block (Figure~\ref{case-1}A1) and a circle of small green blocks (Figure~\ref{case-1}A2) caught his attention.
These green blocks indicated some groups of addresses had traded heavily in NFTs within a relatively short time, which was a sign of wash trading.
Additionally, the areas around these green blocks in the gradient background were much darker than the rest, indicating that these suspicious transactions might have an impact on the price of this collection.

\textbf{A Big Group of Colluding Addresses.}
To further confirm the NFTs involved in wash trading and the trading patterns, U1 first selected the area of the big green block shown in Figure~\ref{case-1}A1 with the circular brush, and then the stacked area chart showed the number of NFTs held by 11 selected addresses at the group level from January 11, 2022, to June 17, 2022 (Figure~\ref{case-1}C).
Overall, U1 found that the height of the stacked area chart increased over the whole time period, while most of the time the height stayed the same.
It indicated that these addresses were holding an increasing number of NFTs at the group level, but they were always trading them within the group.
Furthermore, U1 discovered that the colors in the stacked area chart change in a sophisticated way over time, and he wondered how this group of addresses conducted the wash trading.

By brushing and dragging the brushed area on the stacked area chart, he analyzed the detailed NFT flows of multiple time periods with the flow-based chart, as shown in Figure~\ref{case-1}(D1-D6).
During the period of Figure~\ref{case-1}D1, U1 found four obvious flow graphs of address 7,8,9,10, and each of them kept the height of one or two NFTs. 
It implied that each address still held one or two NFTs during the heavily traded process, which was exactly feature two of wash trading, trading NFTs without taking market risk.
By observing the trade paths of each unique NFT, U1 discovered that each NFT had many repeated transaction patterns, i.e., each NFT was traded in the exact same order.
Then, he deduced that these addresses are trading robots that executed the wash trading automatically based on pre-programmed procedures.
Following that, in Figure~\ref{case-1}D2, U1 found that the flow graphs of these four addresses disappeared while the flow graphs of another four addresses appeared and were connected with dotted lines, indicating that these four addresses directly transferred all NFTs to address 3,4,5,6.
U1 guessed that wash traders created four new addresses to conduct the wash trading to avoid detection.
Similarly, U1 found that this group of addresses heavily traded NFTs in various ways after that period.
For example, they sold all NFTs to only addresses 4 and 5 (Figure~\ref{case-1}D3), and then address 4 sent all NFTs back to address 6 (Figure~\ref{case-1}D4).
In Figure~\ref{case-1}D5, all NFTs were sent to six addresses separately and then traded within two pairs of addresses (Figure~\ref{case-1}D6).
According to the findings above, U1 believed that it was a sophisticated wash trading activity by a big group of colluding addresses, and that they were attempting to create fake transactions that seemed normal with various trading patterns.

\textbf{Simultaneous Appearance of Wash Trading.}
U1 then wanted to investigate the area in Figure~\ref{case-1}A2, where many small green blocks were distributed on a ring area in dark blue.
The majority of the green blocks contained numerous transactions between two addresses, which U1 can roughly attribute to wash trading of address pairs according to their suspicious scores presented by inner paths. 
He changed the background to show the monthly trading volume (Figure~\ref{case-1}B), and found that a light blue ring appeared on the outer area of these green blocks (Figure~\ref{case-1}B1), which was in dark blue when the background displayed the price (Figure~\ref{case-1}A2).
It indicated that after the appearance of these green blocks, the price of this collection rapidly increased while trading volume abruptly decreased.
U1 believed that most transactions during this time were high-priced and created by wash traders, so he brushed these green blocks to check the details.
After brushing the areas in Figure~\ref{case-1}(B2-B4), U1 observed several wash trading activities on a small scale, involving a small number of addresses and NFTs, as shown in Figure~\ref{case-1}(E1-E3).
Specifically, in Figure~\ref{case-1}E1, wash trading was just between two addresses by heavily trading two NFTs.
In Figure~\ref{case-1}E2, two addresses first traded one NFT to each other many times before transferring it to another two addresses to perform wash trading.
From Figure~\ref{case-1}E3, the second address traded NFTs with others alternatively, which might be the main address of this group according to U1.

Overall, after using \techName{}, U1 believed that the \textit{Meebits} collection was filled with various wash trading activities, and wash trading had impacted the price and trading volume of this collection.
U1 said that gaining these insights would have taken much longer without using \techName{}.


\subsection{Case 2: Wash Trading Enhanced by Trading Rewards but Discouraged by Royalties}
U2 is an NFT investor with over six months of experience trading NFT in NFT markets, and  also a product manager of NFT projects.
During the user interview with U2, we first introduced a usage scenario for the \textit{Meebits} collection and then asked U2 to analyze the other three collections (i.e., \textit{BAYC}, \textit{Azuki}, and \textit{Loot}) by using \techName{}.

\begin{figure*}[htb]
  \centering
  \includegraphics[width=1\linewidth]{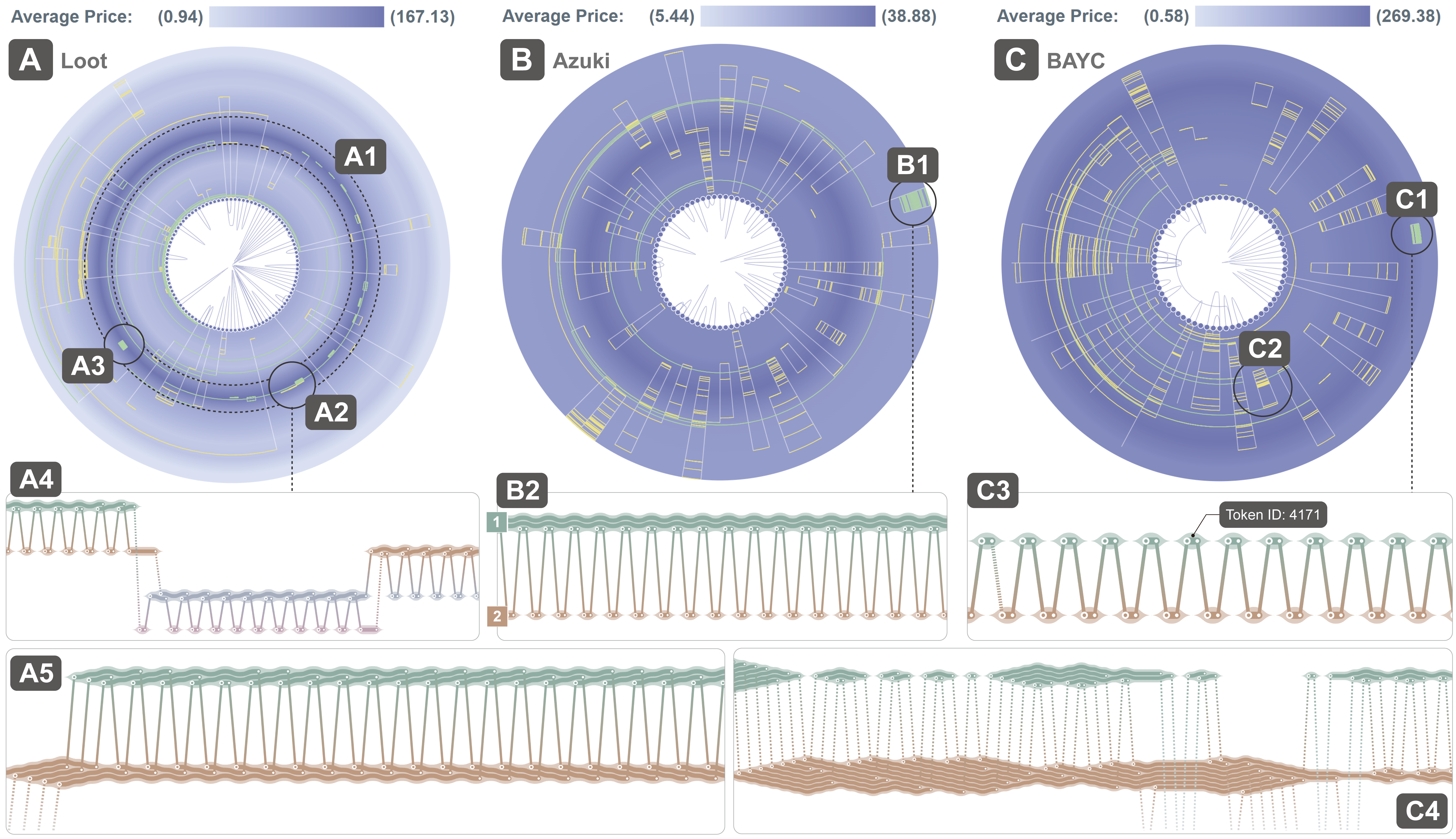}
  \caption{\label{case-2}
            \techName{} shows the wash trading activities in three different NFT collections. (A), (B), and (C) show the \disk{} of \textit{Loot}, \textit{Azuki}, and \textit{BAYC}. (A1) shows a dark blue ring containing many green blocks in the \disk{} of \textit{Loot}. After brushing green blocks in (A2) and (A3), the \flow{} is shown in (A4) and (A5). (B2) shows the NFT flows of the area in (B1). (C3) and (C4) show the \flow{} of the green block in (C1) and the yellow block in (C2). 
           }
    \Description{Figure 6. fully described in the text.}
\end{figure*}

\textbf{Trading Rewards Promoted Wash Trading.}
U2 first analyzed the wash trading activities in the \textit{Loot} collection, and its \disk{} is shown in Figure~\ref{case-2}A.
She found that there were several green blocks on a dark blue ring (Figure~\ref{case-2}A1), which reminded her of a similar pattern she had noticed while watching the usage scenario on the \textit{Meebits} collection.
This pattern indicated that some small groups of addresses had a high volume of transactions during the same time period, and the average price was obviously higher than in other periods.
U2 brushed these green blocks and observed the \flow{} to see if they were wash trading activities.
In Figure~\ref{case-2}A2, she discovered a group with four addresses and saw the detailed NFT flows between them in the \flow{}, as shown in Figure~\ref{case-2}A4.
She brushed the green block in Figure~\ref{case-2}A3 and found there were only two flow graphs in the flow-based chart (Figure~\ref{case-2}A5), where some dotted lines connected from the bottom border to the brown flow graph and all trade paths were between the two flow graphs with little change in height over time.
This indicated that the brown address obtained some NFTs from other addresses and traded them with the green one without taking any market risk.
U2 found that other green blocks in Figure~\ref{case-2}A1 have similar patterns that represented small scale wash trading activities between two addresses.
Furthermore, U2 discovered that the time periods of wash trading in \textit{Loot} and \textit{Meebits} were both in January 2022, and she speculated that some events might have occurred that resulted in the emergence of these wash trading activities.
Then, she thought about the NFT marketplace, LooksRare\textsuperscript{\ref {looksrare}}, was established in January 2022 and became popular rapidly.
U2 said that LooksRare suggested trading rewards for addresses that traded frequently, which could be one of the factors promoting the simultaneous appearance of wash trading in January 2022.

\textbf{Royalties Discouraged Wash Trading.}
Next, U2 used \techName{} to explore \textit{Azuki} and \textit{BAYC} one by one.
In the \disk{} of \textit{Azuki} (Figure~\ref{case-2}B), U2 discovered many yellow blocks on the Outer Ring and only one green block in Figure~\ref{case-2}B1.
The yellow blocks contained many NFT transfers without money, which could have been caused by some addresses belonging to the same person being used for different functions and can not be considered wash trading.
Then, U2 brushed the only green blocks and saw that there were three NFT trade paths in the flow-based chart (Figure~\ref{case-2}B2), two of which are packed in the green flow graph, and one of which traveled between the two flow graphs.
This indicated that the green address held three NFTs and always traded one of them to the brown address to perform the wash trading.
\textit{BAYC}'s \disk{} (Figure~\ref{case-2}C) was similar to \textit{Azuki}'s, with many yellow blocks and only one green block.
Figure~\ref{case-2}C4 showed the \flow{} after brushing the yellow block in Figure~\ref{case-2}C2, where U2 found two flow graphs connected with many dotted lines and the NFTs held by the two addresses changing irregularly, and she believed that it was not wash trading activities.
By checking the \flow{} of the green block (Figure~\ref{case-2}C1), U2 found that two addresses were heavily trading one NFT whose token ID is 4171.
Overall, there were few wash trading activities in both the two collections.
Since U2 is a NFT product manager, she really wondered why there have not been so many wash trading activities in January 2022, unlike the other two collections, i.e., \textit{Meebits} and \textit{Loot}.
She discovered that trading NFTs of \textit{BAYC} and \textit{Azuki} required a certain amount of money, called royalties, to be paid to the owner of the collection or marketplace, whereas trading the other two collections did not.
In this way, U2 got a rough insight that trading rewards might promote wash trading while royalties restrict it, and said that she is more likely to purchase NFTs with royalties in the future.

\section{User Interview \label{sec:interview}}
We conducted semi-structured user interviews with 14 NFT investors to demonstrate the effectiveness and usability of \techName{}.
\modify{Specifically, we asked the investors to analyze the wash trading activities in real NFT collections with \techName{} and further collected their feedback on \techName{}.
}
This section first introduces the setup of our interviews, like participants' backgrounds, interview procedures, and tasks, and then summarizes the feedback collected from participants.

\subsection{Participants and Apparatus}
We recruited 14 target users (U1-U14) from various NFT communities for the user interviews (4 females, 10 males, $age_{mean}=28$, $age_{sd}=5.45$, with normal vision and no color-blindness).
\modify{
All the participants are experienced cryptocurrency investors with a experience of at least two years, and they have been creating or investing in the NFT market.
The detailed information of our participants is summarized in Table~\ref{participant}.
}
U12-U14 are novices with less than six months of NFT trading experience, and U1-U11 have all traded NFTs for more than six months and are proficient with existing NFT data analysis tools, with U3 being the second expert (E2) that attended our preliminary study.
U1, U4-U6 are the creators of some NFT communities that include thousands of investors, with U1 and U4 being key opinion leaders (KOLs) on Twitter with over ten thousand followers.
U2, U7, and U8 have engaged in the issuance of NFT projects, and U13 is also a professor who studies the digital economy.
Due to the COVID-19 pandemic, our interviews were conducted online via Zoom.
We launched the prototype system of \techName{} on the server and allowed participants to assess it via their own laptops or desktops.
\modify{Each interview lasted about one hour, and we paid a compensation of $\$$15 to 
each participant for their time in our user interviews.}

\begin{table*}[]
\caption{\label{participant}
\modify{
The detailed information of the user interview participants.
All participants have a cryptocurrency investment experience for at least two years,
and their detailed NFT investment experiences are listed in the third column.
}
}
\begin{tabular}{ccccl}
\hline
ID & Gender & Age & NFT Experience & Description\\\hline
    U1    &    Male    &  23   &      13 months &  A creator of an NFT community and a key opinion leader on Twitter.          \\
    U2    &     Female   &  25   &    8 months   &  A product manager for multiple NFT projects.    \\
    U3    &    Male    &   30  &   12 months    &     An NFT investor who is good at using NFT analysis tools.   \\
    U4    &   Female     &   26  &     12 months    &    A creator of an NFT community and a key opinion leader on Twitter.   \\
    U5    &    Male    &   29  &  10 months    &    A creator of an NFT community and a leader of an NFT project.      \\
    U6    &     Male   &  25   &    12 months   &      A creator of an NFT community and a leader of three NFT projects.  \\
    U7    &    Female    &  23   &   7 months     &  An NFT investor engaged in the issuance of NFT projects.      \\
    U8    &     Male   &    27 &    10 months     &   An NFT investor engaged in the issuance of NFT projects.    \\
    U9    &    Male    &   27  &    6 months    &     An NFT investor who is good at using NFT analysis tools.   \\
    U10    &    Male    &  30   &  12 months  &   An NFT investor investing in cryptocurrencies for five years.    \\
    U11   &     Male   &   25  &   7 months        &   An NFT investor investing in cryptocurrencies for two years.  \\
    U12   &      Male  &   28  &      5 months      &  An NFT investor investing in cryptocurrencies for two years.  \\
    U13   &     Male   & 46 &  4 months & A professor whose research focus is digital economy.       \\
    U14    &     Female   &  28 &    5 months & A PhD student with two-year research experience in cryptocurrencies.          \\\hline
\end{tabular}
\end{table*}


\subsection{Tasks}
We collected transactions of four different NFT collections that mentioned in Section~\ref{sec:case} and used them in our user interview.
Specifically, we chose one of them for a tutorial usage scenario and asked participants to analyze the other three collections using \techName{}.
The collections are selected sequentially to ensure that each collection can be analyzed by participants.
During the interviews, the participants were asked to perform the following four tasks sequentially on each collection, which guided them to go through \techName{}’s workflow.
\begin{itemize}
    \item T1. Initialize the visualization by using interactions components to filter out undesired information.
     \item T2. Observe the \disk{} to find suspicious addresses and time periods and brush to select them.
    \item T3. Analyze the NFT flows at the group level by the stacked area chart of the \flow{}.
    \item T4. Brush a period in the stacked area chart and check the detailed NFT flows in the flow-based chart.
\end{itemize}
Participants were allowed to repeat these tasks until they had a thorough understanding of the wash trading activities in this collection.

\subsection{Procedure}
\begin{figure*}[htb]
  \centering
  \includegraphics[width=1\linewidth]{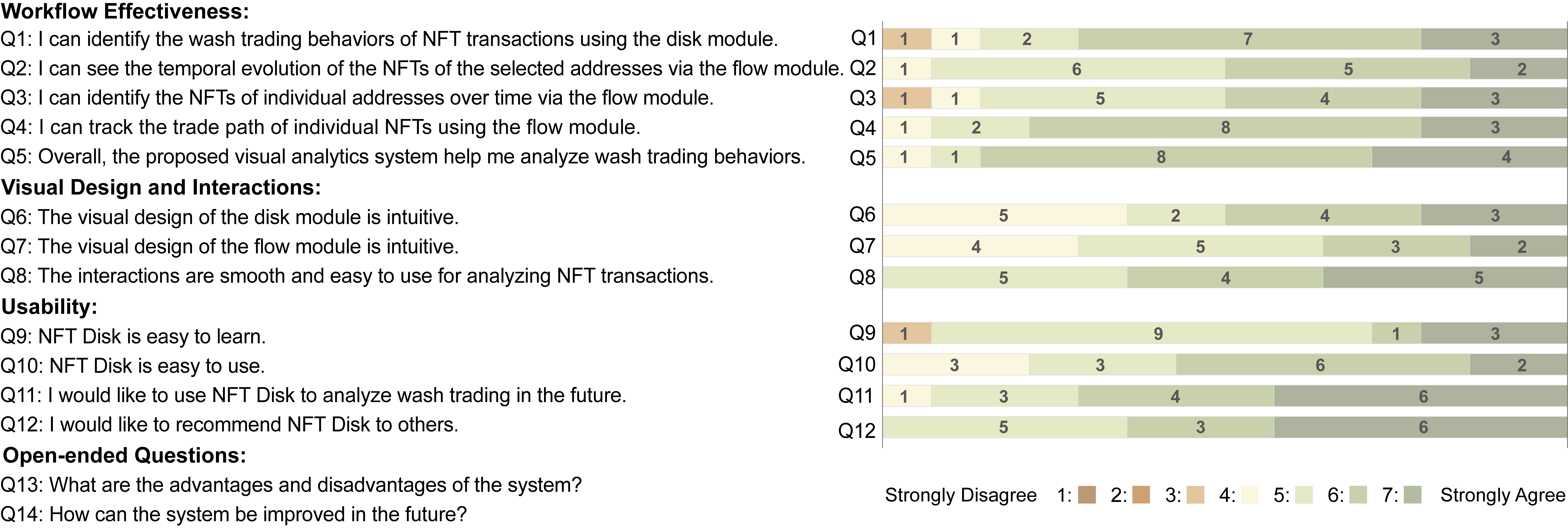}
  \caption{\label{interview}
            The results of the questionnaire in the user interview. Q1-Q12 are close-ended questions designed to evaluate the workflow effectiveness (Q1-Q5), visual design and interactions (Q6-Q8), and usability (Q9-Q12) of \techName{}. Q13-Q14 are open-ended questions to gain the pros and cons of \techName{} from participants.
            Q1-Q12 should be rated by 1-7 representing “strongly disagree” to “strongly agree” on each statement. The number of participants on each rating is represented by a horizontally stacked bar chart.
           }
    \Description[Figure 7. the questionnaire used in the user interview and its results.]{Figure 7. the questionnaire used in the user interview and the results of the closed-ended questions Q1-Q12. Close-ended questions were rated by 1-7 representing “strongly disagree” to “strongly agree” on each statement. Q1-Q5 are designed for workflow effectiveness. Q1: I can identify the wash trading behaviors of NFT transactions using the disk module. Q2: I can see the temporal evolution of the NFTs of the selected addresses via the flow module. Q3: I can identify the NFTs of individual addresses over time via the flow module. Q4: I can track the trade path of individual NFTs using the flow module. Q5: Overall, the proposed visual analytics system help me analyze wash trading behaviors. Q6-Q8 are designed for visual design and interactions. Q6: The visual design of the disk module is intuitive. Q7: The visual design of the flow module is intuitive. Q8: The interactions are smooth and easy to use for analyzing NFT transactions. Q9-Q12 are designed for usability. Q9: NFT Disk is easy to learn. Q10: NFT Disk is easy to use. Q11: I would like to use NFT Disk to analyze wash trading in the future.  Q12: I would like to recommend NFT Disk to others. Q13-Q14 are open-ended questions. Q13: What are the advantages and disadvantages of the system? Q14: How can the system be improved in the future? The results are fully described in the text.}
\end{figure*}

During the interview, we first introduced the background, visual design, interactions, and workflow of \techName{} to the participants.
To show how to use \techName{} to explore the wash trading of NFT collections, we went through an example usage scenario on one of the four collections.
Then, the participants were allowed to explore this collection freely to make themselves familiar with \techName{}.
The tutorial above lasted about 15 minutes.
After that, they were asked to follow the tasks above by using \techName{} to analyze the other three datasets sequentially.
The task phase had no hard time limit and lasted until participants fully understood the wash trading activities in these collections.
In our interview, this phase usually lasted 30 minutes, and their comments and suggestions were recorded.
After the task phase, we invited them to finish a post-study questionnaire with 14 questions (Q1-Q14), as shown in Figure~\ref{interview}.
Q1-Q12 are closed-end questions that should be answered on a 7-point Likert scale and are designed to measure \techName{}'s workflow effectiveness (Q1-Q5), visual design and interactions (Q6-Q8), and usability (Q9-Q12).
\modify{
By following the question design of prior studies~\cite{xie2022roleseer,sun2020dfseer}, Q1-Q5 are designed to assess the workflow effectiveness in terms of different concrete tasks, and Q6-Q8 emphasize the detailed visual designs and interactions.
Q9-Q12 are usability evaluation questions we selected from the PSSUQ (Post-Study System Usability Questionnaire)~\cite{lewis1992psychometric}.
}
Q13-Q14 are open-ended questions that aim to collect participants' feedback on the advantages and possible areas for further improvement of \techName{}.
Overall, each user interview session took about 60 minutes.
\modify{All the data collected from the participants are recorded anonymously with their permission.}

\subsection{Results}
Figure~\ref{interview} summarizes the responses of participants to our post-study questionnaire.
Overall, \techName{} was highly rated by participants.
Most of them agreed that \techName{} was more useful and effective than previous methods for exploring wash trading behaviors in NFT collections.
They stated that they will use it to benefit their NFT investments in the future and recommend it to others.
However, two of the participants gave a rating of less than four because they found it difficult to learn such a complicated visual design and thought that \techName{} was unsuitable for novice investors who had never learnt about the data on blockchain.
The detailed comments and improvements from participants can be summarized as follows:

\textbf{Workflow Effectiveness.}
\modify{
Figure~\ref{interview} shows that most participants provided a positive rating for Q1-Q5, which confirms the the workflow effectiveness of \techName{} in visually identifying and analyzing wash trading activities.
Specifically, U1-U11, who are more experienced NFT investors,
were able to quickly complete the tasks and interpret the patterns shown in the visualizations using their domain knowledge and investment experience, whereas novice participants (U12-U14) needed more time to interpret the visual discoveries and further explain the possible underlying reasons.
U14 scored ``3'' for Q1 and Q3, because it took her some time and effort to fully understand the visual encodings of the \flow{} and \disk{} when analyzing the first NFT collection.
But once she understood the visual designs of \techName{}, she was able to easily finish the exploration and analysis of the remaining two NFT collections.
Overall, 
with the help of \techName{},
all the participants can efficiently identify the wash trading behaviors in the NFT collections and further assess the impact of them.
}

\textbf{Visual Design and Interactions.}
All the ratings on the visual design and interactions are greater than a neutral score (i.e., four), as shown in Figure~\ref{interview}.
U3 liked the design of \disk{} because it can help to quickly overview the suspicious transactions of the entire collection and assess their impact on prices or trading volume.
U6 and U7 phrased that the \flow{} was quite intuitive for showing both the NFT number over time and the trading history of individual NFT, and they emphasize that the \flow{} can be useful for other behavior analysis besides wash trading.
However, five participants gave a rating of four for the design of \disk{} and \flow{} since such visualizations were more sophisticated than simple diagrams such as bar charts or line charts and took time to understand.
\newmod{
These five participants also mentioned that it took some time (about 10 minutes) to learn visual encodings when using \techName{} for the first time, but once they are familiar with the visual designs, they can perform analysis quickly.}
All participants agreed that our interactions were flexible and smooth, and two brush interactions can effectively filter out unwanted information. 
The interactions were highly rated by all participants, but some of them pointed out that the circular brush selected unwanted addresses sometimes.
U8 added, ``Sometimes, the circular brush will selected some undesired middle addresses if the distance between the target addresses is great.''

\textbf{Usability.}
Overall, most participants believed that \techName{} was easy to learn and use (Q9-Q10 of Figure~\ref{interview}).
After learning the usage of \techName{}, the majority of the participants would like to use it in the future and recommend it (Q11-Q12 of Figure~\ref{interview}).
But some participants, like U12 and U13, also stated that it
took them some time for novices like them
to learn and use \techName{}. But once they mastered the usage, they can easily analyze wash trading data and would like to use it in the future.
U6 also mentioned, ``\techName{} is more helpful for people like NFT investors who are
interested in analyzing the behaviors of addresses using blockchain data."
%

\modify{
\textbf{Improvements.}
For the open-ended questions, participants provided some insightful suggestions for future improvements.
U13 suggested that it may be helpful to further add some text descriptions in the user interface of \techName{} to explain visual encoding and data information used by visualizations. Such information can help novice users without a background in data visualization master the usage of \techName{} more easily.
U4 remarked, ``\techName{} can facilitate the analysis of wash trading behaviors in NFT transactions.
But if we want to further analyze the detailed impact of the wash trading behaviors,
we often need to include more information, such as other token transactions and account balances at these addresses.'' Thus, U4 recommended us to further incorporate such information in \techName{}, which is left as future work.
Finally,
U3 pointed out that the address filtering can be more expressive and go
beyond filtering by only the amount of transactions.
Instead, it can enable the filtering by more address attributes such as 
account balance and trading frequency, which has been implemented in the latest prototype of \techName{}.
}

%% file: texfiles/6_coclusion.tex
\section{discussion}

Our case studies and user interviews have demonstrated the effectiveness and usability of \techName{}.
\modify{
In this section, we further discuss the lessons we learned, the specific design considerations for novice users, generalizability and limitations of our approach.
}

\modify{\textbf{Lessons learned.}
During the development of \techName{}, we also learned a few important lessons.
First, when we are trying to detect wash trading in NFT transactions, it is usually not sufficient to focus on the transaction behaviors of only individual addresses, which is the general practice of most existing wash trading detection methods~\cite{das2021understanding, von2022nft}.
Instead, we discovered that wash trading of NFTs often occurred
among a group of addresses,
where the NFTs held by a group of addresses remained the same while the NFTs held by each individual address were changing, such as in Figure~\ref{case-1}D2.
Thus, we have designed the \flow{}, which can intrinsically enable the convenient analysis of transaction behaviors within and across different groups of addresses.
Second, with the help of \techName{},
we also noticed that different addresses in a colluding address group for NFT wash trading may be responsible for different tasks. Some addresses were in charge of collecting NFTs, and some were dedicated to
performing wash trading.
Existing automatic algorithms~\cite{von2022nft,das2021understanding}
can only find addresses performing wash trading but are not able to reveal the addresses that are responsible for collecting NFTs for an NFT wash trading address group, which are both achieved in \techName{} by visualizing their detailed transaction histories. 
Third, not all the wash trading behaviors are ``harmful'' to investors.
When the price of NFTs in a NFT project is suddenly dropping, sometimes the owner of NFT projects may also frequently buy and sell a large number of NFTs in a short period to stabilize the NFT price.
Such kind of ``wash trading'' behaviors by the NFT project owners are often welcomed by investors. 
\techName{} can intuitively reveal such wash trading transactions to investors, assisting them in their decision making for NFT investments.
}


\modify{\textbf{Design considerations for novice users.}
Our target users are NFT investors, but they are not necessarily familiar with data visualizations or even the technical knowledge of blockchain.
During the development of \techName{}, we paid special considerations into such issues and tried to make the overall visual designs and workflow as intuitive as possible. 
First, straightforward visualizations are proposed to represent the complex features of wash trading behaviors. For instance, we proposed the suspicious scores and explicitly encoded them by the height of inner paths, and we used the change in flow height to imply market risk. 
We also translated the professional technical terms in the blockchain data into concepts that general investors are more familiar with, e.g., the technical term ``addresses'' is rephrased as user-friendly concepts like ``buyers'' and ``sellers''.
Second, the overall design and workflow of \techName{} are guided by the well-known visualization mantra of Overview-First, Details-On-Demand~\cite{shneiderman2003eyes}. The \disk{} provides an overview of NFT transactions and reveals the suspicious addresses, and the \flow{} enables an in-depth investigation into these suspicious addresses with three levels of details. Such a workflow can assist even novice investors in easily identifying and analyzing the wash trading of NFT markets in a well-organized manner.
\newmod{
In summary, target users' knowledge background and their familiarity with visualizations should be carefully considered in the development of visual designs. Such design considerations can also apply to other real-world applications beyond the analysis of NFT wash trading.
}
}


\modify{
\textbf{Generalizability.}
%
\techName{} is designed for visual detection of wash trading of NFT transactions. However,
the workflow 
can be used as a reference for analyzing other frauds of cryptocurrency that show abnormal transaction patterns, such as Pump and Dumps~\cite{la2020pump} and money laundering~\cite{jordanoska2021exciting}.
%
%
Also, \techName{} can be easily extended to the wash trading detection of stocks and funds in the traditional investment market~\cite{cao2015detecting}, if the detailed transaction data between investors are available. 
Moreover, \techName{} has the potential to work for the analysis of other abnormal online activities involving different participants. One possible application of \techName{} can be the identification and analysis of political astroturfing on Twitter~\cite{keller2020political,zhang2013online}, where a group of Twitter accounts are manipulated by their employer to distribute disinformation in a coordinated manner. The \disk{} can be helpful for overviewing the coordinated disinformation distribution activities of all the involved Twitter accounts, and the \flow{} can be used to reveal the detailed activity evolution of each account. 
}

\textbf{Limitations.}
\modify{
In \techName{}, we proposed a variety of solutions to enhance its visual scalability, such as the address ordering strategy, and the filtering and brushing interactions,
which make \techName{} work well for the majority of collections.}
\modify{
When there are a huge number of suspicious addresses and transactions that need to be checked, it will be impossible to show all of the addresses simultaneously due to the limited screen space.
More processing like clustering similar addresses may be needed to further enhance the scalability. 
}
Another issue originates from the quality of the data on the blockchain. 
Even though all the transactions that occurred were recorded on the blockchain, there was also some information that is difficult to be identified in the transactions.
For example, some NFT trading tools have emerged in the market to support bulk trading of NFTs, which can hide important information like price and cause some sales of NFTs to be identified as NFT transfers. 
It can be solved by combining the data record on the blockchain with the off-chain data provided by the NFT marketplaces.
\newmod{
Moreover, a quantitative time cost comparison between using \techName{} and other baseline approaches can further demonstrate the effectiveness of our approach in a more convincing way.
However, we did not perform such a time cost comparison,
and the major dilemma is that there are no such existing tools for investors to investigate wash trading activities. The current practice is to simply check the original transaction records, which is obviously less efficient than using our approach. 
}






\section{conclusion}
We propose a top-down workflow for investors to visually detect and analyze wash trading activities in NFT markets, and present \techName{}, an interactive visualization comprised of two novel visualization modules: the \disk{} and the \flow{}.
Specifically, the \disk{} is a radial visualization module with a disk metaphor to overview NFT transactions and help quickly identify potential wash trading behaviors, while the \flow{} is designed to demonstrate detailed NFT flow at multiple levels.
A diverse set of interactions is also included to enhance the workflow. 
For evaluation, we describe two case studies and conduct an in-depth user interview with 14 real NFT investors.
The results demonstrate that \techName{} is useful and effective in exploring wash trading in NFT markets.

\modify{
In future work, we plan to involve automatic approaches to facilitate the initial filtering of suspicious addresses,
reducing the manual efforts of wash detection and further enhancing the effectiveness of \techName{}.
Moreover, we would like to incorporate more information, like the money flow of addresses, to support the identification of addresses involved in wash trading and assess its detailed impact.
}


\begin{acks}
This work was done during Xiaolin Wen's internship at the Singapore Management University (SMU) under the supervision of Dr. Yong Wang. This work was supported by the Singapore Ministry of Education (MOE) Academic Research Fund (AcRF) Tier 1 grant (Grant number:
20-C220-SMU-011), Lee Kong Chian Fellowship awarded to Dr. Yong Wang by SMU, and the National Natural Science Foundation of China (Grant number: 62172289).
We would like to thank the participants in our user interviews and anonymous reviewers for their
feedback.
\end{acks}